\documentclass[pra,aps,twocolumn,epsfig,superscriptaddress,showpacs]{revtex4}
\usepackage{bm}
\usepackage{amsfonts}
\usepackage[dvips]{graphicx}
\usepackage{mathrsfs}
\usepackage[intlimits]{amsmath}
\usepackage[colorlinks=false,dvipdfm]{hyperref}
\begin{document}
\title{All-versus-nothing violation of local realism in the one-dimensional
Ising model}
\author{Dong-Ling Deng}
\author{Jing-Ling Chen}
 \email{chenjl@nankai.edu.cn} \affiliation{Theoretical Physics Division, Chern Institute of
Mathematics, Nankai University, Tianjin 300071, People's Republic of
China}

\date{\today}

\begin{abstract}
We show all-versus-nothing proofs of Bell's theorem in the
one-dimensional transverse-field Ising model, which is one of the
most important exactly solvable models in the field of condensed
matter physics. Since this model can be simulated with nuclear
magnetic resonance, our work might lead to a fresh approach to
experimental test of the Greenberger-Horne-Zeilinger contradiction
between local realism and quantum mechanics.
\end{abstract}


\maketitle

The all-versus-nothing (AVN) violation of local realism
~\cite{1990Mermin}, which was  noted first by
Greenberger-Horne-Zeilinger (GHZ)~\cite{GHZ-TH}, reveals most
strongly the contradiction between the Einstein-Podolsky-Rosen's
(EPR's) local realism (LR)~\cite{A.Einstein} and quantum mechanics.
Unlike the original Bell's theorem in the form of
inequality~\cite{J.S.Bell} that shows inconsistency between LR and
statistical correlations of quantum mechanics, the AVN proof
demonstrates directly a conflict between LR and nonstatistical
predictions of quantum mechanics. Thus the quantum non-locality can
be manifest in a single run of a certain measurement. It opens a new
chapter on the hidden variables
problem~\cite{M.Zukowski-1991,2001Cabello2,2001Cabello,1993Hardy}
and inspires the quantum protocols for secret
sharing~\cite{M.Zukowski-1998} as well as for reducing communication
complexity~\cite{1997Cleve}. The experimental tests have been made
by using the multi-photon
entanglement~\cite{2000Pan,1999Bouwmeester}.

However, most of the previously constructed GHZ contradiction
started from given quantum states but not much concerned the
Hamiltonian of a physical system. Although a given quantum state can
be created theoretically by quantum controllability, the techniques
needed may be very complex even for a few qubits. In contrast, in
condensed matter physics, the dynamical evolution of a physical
system is governed completely by its Hamiltonian and the
corresponding states especially the ground states exist naturally.
What's more, the measurements of muti-photon states are subversive,
i.e., after the test of the GHZ contradiction, the multipartite
entanglement between photons no longer exist. Concerning this
problem, one would ask naturely: Can we find AVN proofs in  physical
systems with specific Hamiltonians?

Actually, besides multi-photons, there are many other interesting
physical systems that may provide platforms for us to construct and
test a GHZ contradiction. For instance, in Ref.~\cite{2009Hu}, an
AVN proof of Bell's theorem was showed in the Kitaev's toric code
model, which is an exactly solvable model and crucial for
fault-tolerant topological quantum computation
(TQC)~\cite{2003Kitaev,2006Kitaev}. Moreover, a possible
experimental scheme using anyonic interferometry~\cite{2008Jiang} to
test the GHZ contradiction in this model was also proposed. This
approach has a predominant advantage compared with the experimental
tests by using multiphoton systems, namely, the measurement employed
in the experimental scheme is nondestructive and can be repeated
without disturbing the ground state. Nevertheless, despite notable
progresses in this direction~\cite{2007Han}, experimental
realization of anyonic interferometry in the Kitaev's toric code
model would still be a challenging task hitherto.

Another attractive alternative concerns the one-dimensional
transverse-field Ising model~\cite{IsingM}, which has a simple and
clear physical picture and can be solved
exactly~\cite{IsingM-ExactS}.  This model consists of many
spin-$1/2$ particles on a chain with the nearest neighboring Ising
interaction. It is often used as a starting model to test new
physical ideas and approaches due to its simplicity and clear
physical pictures. Moreover, this model can be simulated in the
laboratory with present technologies, for instance with nuclear
magnetic resonance (NMR) technologies~\cite{JFZhang2008}, thus might
leading to a fresh approach to experimental test of the GHZ
contradiction.

In this paper, we show AVN violations of LR  in the one-dimensional
transverse-field Ising model by explicitly presenting a set of
commuting operators that when acting on the ground states admit the
GHZ contradiction. We restrict our study on the ground states of
this model and set the external transverse magnetic field to be
zero. 
But we find that this is not necessary. Some excited states are also
useful for the AVN proofs of the GHZ contradiction.

\begin{figure}[tbp]
\includegraphics[width=75mm]{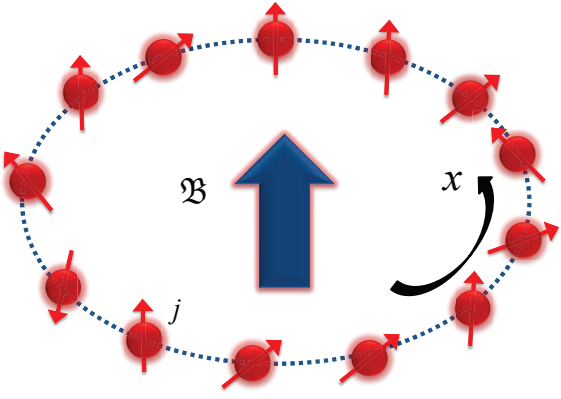}\newline
\caption{(Color online) A illustrative sketch of the one-dimensional
transverse-field Ising model of thirteen spins with periodic
boundary conditions. In this model, all spins are subject to an
external field $h$ along $z$ direction and two arbitrary neighboring
spins interact with each other by the Ising interaction
$\protect\sigma^x_j\protect\sigma^x_{j+1}$. } \label{IsingFig}
\end{figure}

To begin with, let us first briefly recapitulate the essence of the
Ising model. The Hamiltonian of the one-dimensional transverse field
Ising model with periodic boundary conditions reads:
\begin{eqnarray}\label{NqubitHam}
\mathscr{H}_N=-\sum_{j=1}^{N}(\sigma_j^x\sigma_{j+1}^x+\mathscr
{B}\sigma_j^z),\quad\sigma^x_{N+1}=\sigma^x_1,
\end{eqnarray}
where $\sigma^x_j$, $\sigma^z_j$ are the Pauli matrices for the
$j$th spin, $N$ is the number of spins involved in the model and
$\mathscr{B}$ is the transverse field. As illustrated in
Fig.~\ref{IsingFig}, the Hamiltonian describes a chain of spins with
the nearest neighboring Ising interaction along $x$-direction, and
all spins are subject to a transverse magnetic field $\mathscr{B}$
along the $z$-direction. It is well-known that $\mathscr{H}_N$ can
be diagonalized exactly through three transformations, namely, the
Jordan-Wigner transformation~\cite{1928Jordan}, Fourier
transformation, and Bogoliubov transformation. While, for a
rudimental understanding of the ground sates of this model, we only
look at its two limiting cases. If $\mathscr{B}\rightarrow\infty$,
the Ising interaction is neglectable, and all spins are fully
polarized along $z$-direction. If $\mathscr{B}=0$, the Hamiltonian
becomes the classical one-dimensional Ising model. Moreover, it is
easy to prove that the ground state for a finite system in the whole
region $\mathscr{B}>0$ is nondegenerate. While, for $\mathscr{B}=0$,
the ground state is doubly degenerate. Our construction of the AVN
proof is based on one of the doubly degenerate ground states.

%

To present AVN violations of local realism in this model, let us
first focus on the three qubit-case, namely $N=3$. In this case, the
Hamiltonian of this model in the matrix form reads:
\begin{eqnarray}
\mathscr {H}_3=\left(\begin{matrix}
-3\mathscr{B}&0&0&-1&0&-1&-1&0\\
0&-\mathscr{B}&-1&0&-1&0&0&-1\\
0&-1&-\mathscr{B}&0&-1&0&0&-1\\
-1&0&0&\mathscr{B}&0&-1&-1&0\\
0&-1&-1&0&-\mathscr{B}&0&0&-1\\
-1&0&0&-1&0&\mathscr{B}&-1&0\\
-1&0&0&-1&0&-1&\mathscr{B}&0\\
0&-1&-1&0&-1&0&0&3\mathscr{B}
\end{matrix}\right).
\end{eqnarray}
By solving the Hamiltonian $H_3$ directly, one can easily find that
the ground state of this Hamiltonian is:
\begin{eqnarray}\label{3SpinGS}
|{\rm
G}\rangle_3=(\xi_1|000\rangle+|011\rangle+|101\rangle+|110\rangle)/\sqrt{\mathscr
{N}_1},
\end{eqnarray}
where $\xi_1=-1+2\mathscr{B}+2\sqrt{1-\mathscr{B}+\mathscr{B}^2}$
and
$\mathscr{N}_1=3+\xi_1^2$ is the normalization constant. 
Intuitively, when $\mathscr{B}\neq0$, then $\xi_1\neq1$ and it is
hard to construct the set of commuting operators for the GHZ
contradiction. In fact, we have run a computer program hoping to
find such a set of operators that would lead to the GHZ
contradiction when $\mathscr{B}\neq0$. Unfortunately, no valuable
result has been found. Thus, we have to consider the case
$\mathscr{B}=0$. In this case, the ground state of the Hamiltonian
$\mathscr{H}_3$ reduces to $|{\rm
G}(\mathscr{B}=0)\rangle_3=\frac{1}{2}(|000\rangle+|011\rangle+|101\rangle+|110\rangle)$.
Based on the ground state $|{\rm G}(\mathscr{B}=0)\rangle_3$, it is
easy to show the AVN proof of Bell's theorem in the one dimensional
Ising model. For convenience, we should define the following
operators which commute with each other:
$\mathscr{D}^1_3=\sigma_1^y\sigma_2^y\sigma_3^z$,
$\mathscr{D}^2_3=\sigma_1^y\sigma_2^z\sigma_3^y$,
$\mathscr{D}^3_3=\sigma_1^z\sigma_2^y\sigma_3^y$, and
$\mathscr{D}^4_3=\sigma_1^z\sigma_2^z\sigma_3^z$. On the one hand,
one can easily check the following eigenequations for the above four
commuting operators:
\begin{subequations}
\begin{eqnarray}
\sigma_1^y\sigma_2^y\sigma_3^z|{\rm G}(\mathscr{B}=0)\rangle_3&=&-|{\rm G}(\mathscr{B}=0)\rangle_3\label{ThreeQum1},\\
\sigma_1^y\sigma_2^z\sigma_3^y|{\rm G}(\mathscr{B}=0)\rangle_3&=&-|{\rm G}(\mathscr{B}=0)\rangle_3\label{ThreeQum2},\\
\sigma_1^z\sigma_2^y\sigma_3^y|{\rm G}(\mathscr{B}=0)\rangle_3&=&-|{\rm G}(\mathscr{B}=0)\rangle_3\label{ThreeQum3},\\
\sigma_1^z\sigma_2^z\sigma_3^z|{\rm
G}(\mathscr{B}=0)\rangle_3&=&+|{\rm
G}(\mathscr{B}=0)\rangle_3\label{ThreeQum4}.
\end{eqnarray}
\end{subequations}

Suppose there are three observers, each of them having access to one
spin in the model. On spin $j$, the corresponding observer measures
either $\sigma_j^y$ or $\sigma_j^z$, without disturbing the other
spins. The results of these measurements will denote by $m^y_j$ or
$m^z_j$, respectively. Since these results must satisfy the same
functional relations satisfied by the corresponding operator, then
from Eqs. (\ref{ThreeQum1})-(\ref{ThreeQum4}), we can predict that,
if all the operators in Eqs. (\ref{ThreeQum1})-(\ref{ThreeQum4}) are
measured, their results must satisfy
\begin{subequations}
\begin{eqnarray}
m^y_1m^y_2m^z_3&=&-1,\label{ThreeClass1}\\
m^y_1m^z_2m^y_3&=&-1,\label{ThreeClass2}\\
m^z_1m^y_2m^y_3&=&-1,\label{ThreeClass3}\\
m^z_1m^z_2m^z_3&=&+1.\label{ThreeClass4}
\end{eqnarray}
\end{subequations}

On the other hand, note that
Eqs.~(\ref{ThreeClass1})-(\ref{ThreeClass4}) contain only local
operators, therefore the operators in each equation commute and can
all simultaneously have their eigenvalues. Thus, from EPR's
criterion of elements of reality~\cite{A.Einstein}: \textit{``If,
without in any way disturbing a system, we can predict with
certainty (i.e., with probability equal to unity) the value of a
physical quantity, then there exists an element of physical reality
corresponding to this physical quantity"}, we can associate an EPR
element of physical reality to each of the eigenvalues in
Eqs.~(\ref{ThreeClass1})-(\ref{ThreeClass4}). For instance, the
observers on particles $2$ and $3$ measure, respectively and without
disturbing each other, $\sigma_2^y$ and $\sigma_3^z$. If the
multiplier of their results is $1$, then from Eq.(\ref{ThreeClass1})
they can predict with certainty that the result of measuring
$\sigma_1^y$ will be $-1$. Other wise, if the multiplier is $-1$,
then they can predict certainly that the result of measuring
$\sigma_1^y$ will be $1$. Anyhow, we can predict with certainty the
value of quantity $\sigma_1^y$ by measuring other particles'
measurements without disturbing particle $1$, therefore, we can
associate a element of reality to the physical quantity
$\sigma_1^y$. Analogously, we can associate elements of reality to
all the physical quantities in
Eqs.~(\ref{ThreeClass1})-(\ref{ThreeClass4}). Then we can suppose
that this result was somehow predetermined and initially hidden in
the original state of the system. Such predictions with certainty
would lead us to assign values $+1$ or $-1$ to all the observables
in Eqs.~(\ref{ThreeQum1})-(\ref{ThreeQum4}). However, such
assignment cannot be consistent with the rules of quantum mechanics
because the four Eqs.~(\ref{ThreeClass1})-(\ref{ThreeClass4}) cannot
be satisfied simultaneously, since the product of their left-hand
sides is $+1$, while the product of the right-hand sides is $-1$.
Therefore, we conclude that the four predictions of quantum
mechanics given by Eqs.~(\ref{ThreeQum1})-(\ref{ThreeQum4}) cannot
be reproduced by LR. Thus we successfully construct the GHZ
contradiction in the one-dimensional Ising model with three spins.
Note that our construction only explore one of the doubly degenerate
ground states of $\mathscr{H}_3(\mathscr{B}=0)$. In fact, one can
also present the AVN proof by using the other degenerate ground
state: $|{\rm
G}'(\mathscr{B}=0)\rangle_3=\frac{1}{2}(|001\rangle+|010\rangle+|100\rangle+|111\rangle)$.

For four-qubit case, i.e., $N=4$, the ground states of the
Hamiltonian $\mathscr{H}_4$ reads:
\begin{eqnarray}
|{\rm
G}\rangle_4&=&[(\xi_2-\frac{2\sqrt{2}\mathscr{B}(1-\xi_3^2)}{\xi_3})|0000\rangle+
(\mathscr{B}+\frac{\xi_3}{\sqrt{2}})|0011\rangle\nonumber\\
&+&\frac{4\mathscr{B}+2\sqrt{2}\xi_3}{2\sqrt{2}\xi_3}|0101\rangle+(\mathscr{B}+\frac{\xi_3}{\sqrt{2}})|0110\rangle\nonumber\\
&+&(\mathscr{B}+\frac{\xi_3}{\sqrt{2}})|1001\rangle+\frac{4\mathscr{B}+2\sqrt{2}\xi_3}{2\sqrt{2}\xi_3}|1010\rangle\nonumber\\
&+&(\mathscr{B}+\frac{\xi_3}{\sqrt{2}})|1100\rangle+|1111\rangle]/\sqrt{\mathscr{N}_2},
\end{eqnarray}
where $\xi_2=-1+2\mathscr{B}^2+2\sqrt{1+\mathscr{B}^4}$,
$\xi_3=\sqrt{1+\mathscr{B}^2+\sqrt{1+\mathscr{B}^4}}$ and
$\mathscr{N}_2=1+3(\mathscr{B}+\frac{\xi_3}{\sqrt{2}})^2
+\frac{1}{4}(2\mathscr{B}+\sqrt{2}\xi_3)^2+\frac{(4\mathscr{B}+2\sqrt{2}\xi_3)^2}{4\xi_3^2}
+(\xi_2-\frac{2\sqrt{2}\mathscr{B}}{\xi_3}+2\sqrt{2}\mathscr{B}\xi_3)^2$
is the normalization constant. A similar analysis shows that the
ground state with $\mathscr{B}=0$: $|{\rm
G}(\mathscr{B}=0\rangle_4=\frac{1}{2\sqrt{2}}(|0000\rangle+|0011\rangle+|0101\rangle
+|0110\rangle+|1001\rangle+|1010\rangle+|1100\rangle$, also leads to
an AVN proof of the GHZ contradiction. Actually, one can construct
the following four operators:
$\mathscr{D}^1_4=\sigma_1^z\sigma_2^y\sigma_3^y\sigma_4^z$,
$\mathscr{D}^2_4=\sigma_1^y\sigma_2^z\sigma_3^y\sigma_4^z$,
$\mathscr{D}^3_4=\sigma_1^y\sigma_2^y\sigma_3^z\sigma_4^z$ and
$\mathscr{D}^4_4=\sigma_1^z\sigma_2^z\sigma_3^z\sigma_4^z$,  and
then obtain the following corresponding eigenequations:
\begin{subequations}
\begin{eqnarray}
\sigma_1^z\sigma_2^y\sigma_3^y\sigma_4^z|{\rm G}(\mathscr{B}=0)\rangle_4&=&-|{\rm G}(\mathscr{B}=0)\rangle_4\label{FourQum1},\\
\sigma_1^y\sigma_2^z\sigma_3^y\sigma_4^z|{\rm G}(\mathscr{B}=0)\rangle_4&=&-|{\rm G}(\mathscr{B}=0)\rangle_4\label{FourQum2},\\
\sigma_1^y\sigma_2^y\sigma_3^z\sigma_4^z|{\rm G}(\mathscr{B}=0)\rangle_4&=&-|{\rm G}(\mathscr{B}=0)\rangle_4\label{FourQum3},\\
\sigma_1^z\sigma_2^z\sigma_3^z\sigma_4^z|{\rm
G}(\mathscr{B}=0)\rangle_4&=&+|{\rm
G}(\mathscr{B}=0)\rangle_4\label{FourQum4}.
\end{eqnarray}
\end{subequations}
Consequently, the results of these independent measurements on the
four spins satisfy:
\begin{subequations}
\begin{eqnarray}
m^z_1m^y_2m^y_3m^z_4&=&-1,\label{FourClass1}\\
m^y_1m^z_2m^y_3m^z_4&=&-1,\label{FourClass2}\\
m^y_1m^y_2m^z_3m^z_4&=&-1,\label{FourClass3}\\
m^z_1m^z_2m^z_3m^z_4&=&+1.\label{FourClass4}
\end{eqnarray}
\end{subequations}
Based on Eqs~(\ref{FourClass1})-(\ref{FourClass4}),  the assignment
of predetermined values to all the observables in
Eqs~(\ref{FourQum1})-(\ref{FourQum4}) is also impossible because
such assignment would lead to the inconsistence in
Eqs~(\ref{FourClass1})-(\ref{FourClass4}). This completes the AVN
proof of the GHZ contradiction in the one-dimensional Ising model
with four qubits.

Our methodology can be generalized straightforwardly to the
one-dimensional Ising model involving more spins, even though we
have not explored this direction in detail. Briefly, the four
commuting operators can be constructed as
$\mathscr{D}_N^1=\sigma_1^z\sigma_2^y\sigma_3^y\sigma_4^z\cdots\sigma_N^z$,
$\mathscr{D}_N^2=\sigma_1^y\sigma_2^z\sigma_3^y\sigma_4^z\cdots\sigma_N^z$,
$\mathscr{D}_N^3=\sigma_1^y\sigma_2^y\sigma_3^z\sigma_4^z\cdots\sigma_N^z$,
and
$\mathscr{D}_N^4=\sigma_1^z\sigma_2^z\sigma_3^z\sigma_4^z\cdots\sigma_N^z$.
In fact, the validity of the generalization can also be guaranteed
by the evidence that the ground state $|{\rm
G}(\mathscr{B}=0)\rangle_N$ of $\mathscr{H}_N$ is equivalent to the
maximally entangled state in the GHZ form
$|\Psi\rangle_N=\frac{1}{\sqrt{2}}(|00\cdots0\rangle+|11\cdots1\rangle)$
due to local unitary transformations~\cite{2008Gu}.

It is also worthwhile to point out that in above analysis, only one
of the ground states with $\mathscr{B}=0$ are considered. However,
this is not a necessary condition for the AVN proofs in the model.
Actually, one can construct the GHZ contradiction by using various
excited states. For instance, in the four-spin case with
$\mathscr{B}=0$, the first excited states are degenerate and have an
eigenenergy $\mathscr{E}_4^1=0$. One of these first excited states
takes the form:
$|\psi\rangle^{fe}_4=\frac{1}{\sqrt{2}}(-|0000\rangle+|1111\rangle)$.
Then, it is easy to construct four commuting operators
$\mathscr{D}'^{1}_4=\sigma_1^x\sigma_2^x\sigma_3^y\sigma_4^y$,
$\mathscr{D}'^{2}_4=\sigma_1^x\sigma_2^y\sigma_3^x\sigma_4^y$,
$\mathscr{D}'^{3}_4=\sigma_1^x\sigma_2^y\sigma_3^y\sigma_4^x$, and
$\mathscr{D}'^{4}_4=\sigma_1^x\sigma_2^x\sigma_3^x\sigma_4^x$ to
present the GHZ contradiction. In a way, it turns out that the GHZ
contradiction is very common in the one-dimensional Ising model both
for the ground and excited states. We use the ground states to show
the AVN proofs only for the sake of simplicity and convenience.

Recent experiments designed to test the GHZ contradiction are mainly
based on the photons~\cite{2000Pan,1999Bouwmeester}. However, it has
been argued that all these experiments suffer from the
detector-efficiency ``loophole'' in a similar manner as the Bell
inequality~\cite{1999Larsson}. Fortunately, there exist many other
options. For example, the rapid development in the field of Nuclear
Magnetic Resonance Quantum Information Processing (NMR-QIP) has
definitely proven the feasibility and importance of NMR technologies
in quantum information science (for recent reviews, please see
Ref.~\cite{2004Vandersypen} and references there in). In fact, after
more than fifty years of development, NMR has earned an unique
position to perform complex experiments and many physical models,
including the one-dimensional transverse-field Ising model, can be
simulated using the NMR
technologies~\cite{2005Peng,2005Negrevergne}. Recently, NMR
experiments with as many as twelve qubits have been
reported~\cite{2005Lee}. Because of the fundamental importance of
the GHZ paradox in understanding multipartite nonlocality and in
building some quantum information protocols, further experimental
tests of the GHZ paradox might be widely welcomed. Based on the
discussion above, an experiment utilizing NMR technologies to make
such a test in the one-dimensional transverse-field Ising model is
an appealing option. Actually, in Ref.~\cite{2009Deng}, a similar
experiment was proposed to investigate the genuine multipartite
entanglement (GME) and genuine multipartite nonlocality (GMNL)
problem. We suggest that the two experiment might be incorporated to
a single one. Namely, we first simulate the one-dimensional
transverse-field Ising model using the NMR technologies, and then
perform the measurements according to the above discussion to test
the GHZ contradiction and according to Ref.~\cite{2009Deng} to
investigate the GME and GMNL problems.


In summary, we show AVN proofs of Bell's theorem in the
one-dimensional transverse-field Ising model. We focus our attention
on the ground states of this model. While we also find interestingly
that some excited states also admit the AVN proofs of the GHZ
contradiction. It is noteworthy to mention that our approach
concerns the Hamiltonian of a physics system. Thus it might be
useful for future works on the relation between quantum nonlocality
and the dynamical evolution of a physical system. Finally, we
suggest an experiment utilizing NMR technologies to test the GHZ
contradiction in the one-dimensional transverse-field Ising model.
Since the GHZ paradox is closely related to the foundations of
quantum mechanics and is an important primitive for building quantum
information-theoretic protocols that decrease the communication
complexity~\cite{1997Cleve}, such an experiment would be widely
welcomed.


Dong-Ling Deng gratefully acknowledge Prof. Shi-Jian Gu for
illumination and discussion. This work was supported in part by NSF
of China (Grant No. 10975075), Program for New Century Excellent
Talents in University, and the Project-sponsored by SRF for ROCS,
SEM.

\end{document}